\documentclass[preprint2]{aastex63}

\usepackage{graphicx}
\usepackage[suffix=]{epstopdf}
\usepackage{natbib}
\usepackage{amsmath}
\usepackage{url}
\usepackage{xspace}
\usepackage{fontawesome}
\usepackage{blindtext} 
\usepackage[hyphens]{url} 

\providecommand{\e}[1]{\ensuremath{\times 10^{#1}}}

\newcommand{\Kepler}{\textsl{Kepler}\xspace}
\newcommand{\TESS}{\textsl{TESS}\xspace}

\begin{document}

\title{Llamaradas Estelares: Modeling the Morphology of White-Light Flares}

\correspondingauthor{Guadalupe Tovar Mendoza}
\email{tovarg@uw.edu}

\author{Guadalupe Tovar Mendoza}
\affiliation{Astronomy Department, University of Washington, Box 351580, Seattle, WA 98195, USA}

\affiliation{Astrobiology Program, University of Washington, Box 351580, Seattle, WA 98195, USA}

\author{James R. A. Davenport}
\affiliation{Astronomy Department, University of Washington, Box 351580, Seattle, WA 98195, USA}

\author{Eric Agol}
\affiliation{Astronomy Department, University of Washington, Box 351580, Seattle, WA 98195, USA}

\affiliation{NASA NExSS Virtual Planetary Laboratory, University of Washington, Box 351580, Seattle, WA 98195, USA}

\author{James A. G. Jackman}
\affiliation{School of Earth and Space Exploration, Arizona State University, Tempe, AZ 85287, USA}

\author{Suzanne L. Hawley}
\affiliation{Astronomy Department, University of Washington, Box 351580, Seattle, WA 98195, USA}

%
\begin{abstract}
Stellar variability is a limiting factor for planet detection and characterization, particularly around active M-type stars. Here we revisit one of the most active stars from the \Kepler mission, the M4 star GJ 1243, and use a sample of 414 flare events from 11 months of 1-minute cadence light curves to study the empirical morphology of white-light stellar flares. We use a Gaussian process detrending technique to account for the underlying starspots. We present an improved analytic, continuous flare template that is generated by stacking the flares onto a scaled time and amplitude and uses a Markov Chain Monte Carlo analysis to fit the model. Our model is defined using classical flare events, but can also be used to model complex, multi-peaked flare events. We demonstrate the utility of our model using \TESS data at the 10-minute, 2-minute and 20-second cadence modes. Our new flare model code is made publicly available on GitHub \faGithub\href{https://github.com/lupitatovar/Llamaradas-Estelares}.

\end{abstract}

\section{Introduction}

Stellar flares are energetic events that occur on the surface of stars and are a result of the reconnection of magnetic field lines \citep{Benz2008FlarePhysics}. They are believed to share a common underlying physical formation mechanism and have been observed on all types of main sequence stars that have outer convection envelopes \citep{Pettersen1989ACharacteristics}. For instance, there is evidence of flaring on low mass stars \citep[e.g.][]{Lacy1976UVDATA,Pazzani1981TimeStars,Doyle1990AGeminorum,Panagi1995U-band65}, RS CVn stars \citep[e.g.][]{Osten1999ExtremeMegaseconds}, and on the Sun \citep[e.g.][]{Pearce1990SympatheticFlaring}. On the Sun, we see how the variety of magnetic activity ranging from large scale surface events such as spots, coronal mass ejections (CMEs), and flares gives rise to significant photometric variations \citep{Carrington1859Description1859}. From the Sun, we expect flares to occur near active regions on a star \citep{Benz2010PhysicalObjects}. However not all stars follow this behavior; the M-type star GJ 1243 is an example of a star that has flare events happening all over the surface of the star with no significant correlation to the starspot phase \citep{Hawley2014KeplerDwarfs}. Several studies have argued this is because the spot coverage on young active stars is 80\% or greater \citep[]{Gully-Santiago2017PlacingDiagram,Feinstein2020FlareData}. Meanwhile, for fully convective stars it has been shown that flares may occur at very high latitudes because magnetic fields are emerging close to the stellar rotational poles \citep{Ilin2021GiantLatitudes}. Since the polar regions are always visible, unless the inclination is 90 degrees, we can observe the flare events. The ubiquity of flares present among low mass stars motivates further study of their flare frequencies and morphologies.

Stellar magnetic activity has long influenced our ability to detect and characterize extrasolar planetary systems. For instance, in the case of the flare star AU Mic, a flare event occurred at the same time the planet AU Mic b was transiting \citep{Plavchan2020AMicroscopii}. The flare was masked out, which caused greater uncertainty in the transit ingress/egress profile. In addition to detectability, many teams have shown that strong magnetic activity (i.e flares and CMEs) can affect planetary atmospheres  \citep[e.g.][]{Segura2010TheDwarf,Vida2016InvestigatingPegasi,Tilley2017ModelingPLANET} and thus influence potential habitability by causing runaway greenhouse effects  \citep{Shields2016TheStars}, atmospheric erosion \citep{Lammer2007CoronalZones} and hydrodynamic escape of atmospheres \citep{Luger2015HabitableDwarfs}. By better understanding the temporal evolution or light curve morphology of flares on active M-type stars we can help improve exoplanet detection and characterization \citep{Gilbert2021NoData}. 

The \Kepler space telescope has provided long duration, high-precision, optical light curves that are advantageous for studying stellar variability phenomena \citep{Borucki2010KeplerResults}. Many catalogs of flares have been created from the \Kepler data \citep[e.g.][]{Hawley2014KeplerDwarfs,Davenport2016Flares,Martinez2019AASAS-SN}. The catalogs have been useful tools to aid in our ability to understand and model these stellar energetic events.

While it is generally understood that flares share a common physical origin \citep[e.g.][]{Benz2008FlarePhysics}, there are many different parameterizations that have been used to describe what we see during flare events. Previous studies modeled flares using single exponential profiles, fast rise exponential decay (FRED) profiles, or combinations of a Gaussian plus an exponential \citep[e.g.][]{Walkowicz2011White-lightData,Loyd2014FLUCTUATIONSOBSERVATIONS}. However, these models ignore the two-phase cooling decay that is typically observed during flare events \citep[e.g.][]{Andrews1965FlareStars,Hilton2011TheRate,Davenport201411243}. More recently, many have studied the morphology of flares from white-light flare profiles that have impulsive and gradual phases \citep{Kowalski2013Time-resolvedSpectra}, as well as possible quasi-periodic oscillations during flare events \citep[e.g.][]{Pugh2015ASUPERFLARE}. In addition, higher cadence observations have resolved the flare peaks and found they roll over, emphasizing the need for a continuous model that does not have break points between the rise and decay phases \citep[e.g.][]{Kowalski2016M,Jackman2018Ground-basedNGTS,Jackman2019DetectionSurvey,Howard2021NoFlares}. 

\citet[][hereafter D14]{Davenport201411243} used \Kepler short cadence data from data release 23 (DR23) of GJ 1243 to understand the characteristics of flare light curves and found that when many flares are averaged together, a median flare template can be generated. This has proved to be very useful for modeling flare light curves from a variety of surveys \citep[e.g.][]{Schmidt2019TheASAS-SN} and has helped with modeling transits in the presence of flares \citep[e.g.][]{Luger2017ATRAPPIST-1}. However, this flare template has some major limitations. The D14 model used a piece-wise function to parameterize the flare shape, which causes a discontinuity at the peak of a given flare event. The model used the peak time and amplitude as two of the parameters, which are very sensitive to scaling effects. Finally, this model used a local smoothing function to detrend the underlying starspots around the flares. This approach is not as robust as new statistical methods, such as Gaussian process regression \citep{Rasmussen2006RasmussenLearning}, which also provides the uncertainties of the starspot profile.

In this paper, we derive an updated analytic and continuous flare model to parameterize white-light flare events that addresses the limitations of previous models. We start by considering the model introduced in \citet{Jackman2018Ground-basedNGTS}, which uses the parameterization from \citet{Gryciuk2017FlareCurves} and provides a template for modeling flares. This model is continuous but it is only derived from a small number of individual flares. To address the small number of flares, we use a vetted version of the D14 flare catalog to derive an updated template. Thanks to improvements in \Kepler light curve processing, updates to statistical techniques that allow us to detrend starspots, new parameterizations, and a newly vetted dataset, we are now able to address the limitations of the D14 flare template. We present the details of the updated flare model below and also make the code readily available on GitHub \faGithub\href{https://github.com/lupitatovar/Llamaradas-Estelares}.

 The outline of the paper is as follows: In \S\ref{sec:sample}, we revisit the GJ 1243 flare sample from \Kepler data. In \S\ref{sec:gp} we describe starspot detrending using Gaussian process regression. Using the flare sample and starspot detrending, we reproduce the original flare template in D14. In \S\ref{sec:flaretemp}, we introduce a new analytic flare model adapted from \citet{Jackman2018Ground-basedNGTS} and compare to other analytical models. Then in \S\ref{sec:stacking} we present a new method for constructing the flare template that further addresses some of the limitations present in D14. In \S\ref{sec:fit}, we explain the model fitting procedure and analysis. Next, \S\ref{sec:apps} explores various applications of this new model. We conclude with a discussion of the implications of our study, and the promising future for stellar activity studies that combine \Kepler and \TESS data in \S\ref{sec:discuss}.
 
\begin{figure*}[htbp!]
    \centerline{\includegraphics[width=7in]{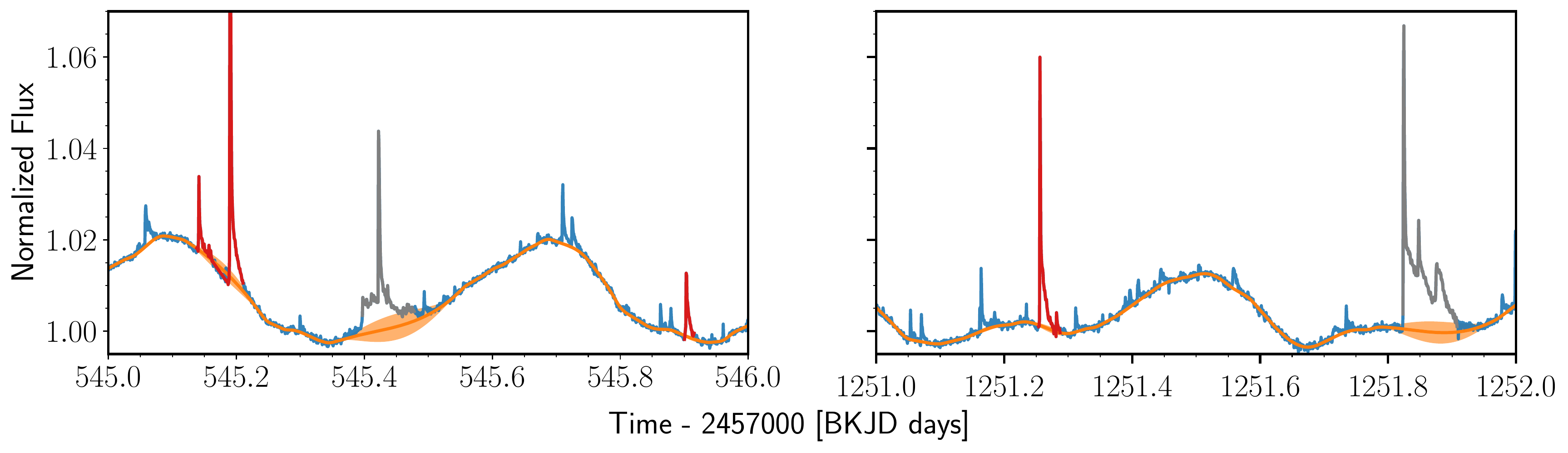}}
    \caption{Two days of GJ 1243 \Kepler 1-minute observations are shown in blue. The identified classical flares (red), the GP mean (orange) and variance are all overlayed. \textbf{Left}: We identified three classical flares during this window. \textbf{Right}: We identified two classical flares that were used to derive the flare model. Note there are two complex flares (grey), one in each panel, but these were not used to construct the flare template. To accurately model the underlying starspot modulations we masked out all flares which allowed the GP to model the starspot variability without being skewed by frequent flaring events. Therefore, we expect the GP to have the highest variance in areas where flares are occurring. }
    \label{GP}
    \centering
\end{figure*}

\section{Defining the Flare Sample}
\label{sec:sample}

For this study, we revisit GJ 1243 (KIC 9726699), one of the benchmark stars for space-based flare studies \citep{Hawley2014KeplerDwarfs}. This dM4e star is the most active flare star in the \Kepler field. The high level of activity observed on the star is directly correlated to its young age of 30-50 Myr \citep{Silverberg2016Kepler1243}. GJ 1243 has a luminosity of log $L_{Kep} = 30.68 \pm 0.04$ erg s$^{-1}$ in the \Kepler bandpass \citep{Davenport2020101243}, an estimated effective temperature of 2,661 K and an estimated mass of 0.094 $M_{\odot}$  \citep{NASAExoplanetArchive2016KeplerTable}. We used 11 months of the {\tt PDCSAP\_FLUX} \citep{Smith2012KeplerCorrection}, 1-minute short cadence light curves from \Kepler Data Release 25\footnote{\url{https://archive.stsci.edu/kepler/release_notes/release_notes25/KSCI-19065-002DRN25.pdf}}, and required the {\tt Quality} flag be set to 0 to minimize the number of errors from spacecraft events. The \Kepler short cadence data was released in months, as opposed to the long cadence data that was released in quarters (1 quarter = 3 months). Therefore, we stitched 11 months of data together in order to create one light curve. We accounted for the quarterly discontinuities by taking the median of the fluxes across each of the months of data and then normalized the light curve by dividing by the total median. 

We use the GJ 1243 flare sample from D14 to study the morphology of white-light flares. To create the sample, D14 developed an IDL tool, Flares By EYE (FBEYE), that ran a smoothing and auto-finding algorithm to identify candidate flares. Users would then manually analyze the light curve to verify and classify flares. The final sample contained 6107 flare events, which is the largest flare sample for a single star to date. 

Since the D14 study, there has been a new \Kepler data release (DR25). DR25 included improvements in light curve processing, which altered the classifications of previously identified flare events. For instance, some flare events in DR23 that were used in D14 now appear to be within the 3$\sigma$ noise limit and/or now appear to be complex events in DR25. We manually inspected all 885 classical flares defined in D14 with the new DR25 light curves. The new DR25 data had 379 flares that were re-classified as complex flare events, as well as smaller flare events that were within the noise limit. These flares were removed from our sample since, as in D14, we are only using classical (single-peak) flare events to derive the flare template. Furthermore, we only used classical flares that had a total duration of at least 20 minutes and omitted any flares whose duration was longer than 75 minutes, since these flares have a higher likelihood of being complex events as found in D14. This yielded a total of 414 classical flare events that were used to construct the flare template compared to 885 flares used in D14. The main reason for the discrepancy has to do with the different data releases used for each study. The new vetted flare list provides a cleaner sample of classical flares that is then used to derive our updated flare template.

\section{Modeling starspot variability}
\label{sec:gp}

In the GJ 1243 light curve, we observe modulations that indicate the presence of two primary groups of long-lived starspots \citep{Davenport2020101243}. It has been shown that complex spot patterns can create non-sinusoidal variations as they rotate in and out of view \citep{Angus2018InferringProcesses}. The evolution of such active regions, combined with differential rotation on the star's surface, can create quasi-periodic signals \citep{Dumusque2011AstronomyRadial-velocities}. Therefore, a strictly periodic model is not a robust or realistic model to use to account for the time series variations. Instead, we need a model that is flexible enough to capture the evolving quasi-periodic behavior that is present.

\subsection{Gaussian Process Regression}
\label{sec:gp regression}
Gaussian processes (GPs) are powerful models that allow us to make predictions about our data even when we do not know the functional form of the model. GPs fit the correlation between points and are defined by a mean function and a covariance matrix \citep{Rasmussen2006RasmussenLearning}. In astrophysics, GPs have been used as a model for stochastic variability in light curves of stars and to model instrumental systematics \citep[e.g.][]{Kipping2012AnStars,Haywood2014PlanetsSystem,  Barclay2015RADIAL,Angus2018InferringProcesses, Barros2020ImprovingVariability}. In the case of stellar variability, GP kernels can be defined to accurately model the photometric variability and temporal evolution of starspot groups. Here we use a GP to improve the detrending of GJ 1243's starspots. In addition, GPs provide the variance of the modulation, which allows us to include the uncertainty resulting from this starspot detrending into our flare model. We can also use the linear component of the GP to account for the monthly variations in the data. 

D14 used a local spot detrending technique to subtract the starspot features around each flare event. Their approach used a custom smoothing function to smooth the light curve and any data that was more than 1$\sigma$ away from the defined boxcar kernel was removed. The resulting light curve was then fit using a cubic spline and the starspot curve was subtracted from the original light curve. While this approach allows for a quick way of identifying flares in the presence of spots, the resulting starspot model can still have contaminant flares present causing us to miss the curvature of the starspot modulation happening under the flare. This is important to highlight because the starspot model is what gets subtracted from the original light curve, that then is used to identify the flares that are used to construct the flare template. Therefore, if the underlying starspot variability is not accurately modeled there is no quantitative uncertainty that can be included into the flare model. 

Here we use a GP for the starspot detrending. The GP was applied to each of the individual 11 months of data to model the underlying variability caused by starspots. Specifically, we use the stochastically-driven, damped simple harmonic oscillator described in \citet{Foreman-Mackey2018ScalableCelerite} as the kernel to model the variability. In general terms, the kernel is the equation that defines the correlation between the given points and is chosen by the user to then define the covariance matrix. We note that newer kernels such as those described in \citet{Gordon2020ACharacterization} include additional terms that account for various noise components which are particularly useful for improving measurements of stellar rotation or transit parameters. However for our case, the simple harmonic oscillator kernel is suitable for describing the correlated noise since the shape of the starspot evolves over timescales much longer than flares and thus the choice of kernel should not have a significant impact on our results. The initial parameters used for the GP follow those described in \citet{Foreman-Mackey2018ScalableCelerite} where, Q= 0.01 and $\omega = \frac{2\pi} {P_{rot}}$. We use a rotation period of 0.59 days, which was measured from light curve modulations of the star spots \citep{Savanov2011Stellar1243} and again with ground based data by \citet{Irwin2011ONSURVEY}. The kernel parameters were optimized by maximizing the likelihood over each month of \Kepler data.

To ensure the GP was not skewed by the frequent flaring events we masked out all of the flares (classical and complex) present in the data. To account for any flare events that might have been cut off (i.e flares with long decay phases), we added a buffer of 0.25 x $t_{flare}$ to the start time and 0.5 x $t_{flare}$ to the stop times of each flare defined in D14, where $t_{flare}$ refers to the duration of the flare event. The result was a mean GP model describing the starspot variability for GJ 1243 with the corresponding variance of the model as highlighted in Figure \ref{GP}. The variance envelopes (i.e areas where the model has the highest uncertainty) correspond to places where flare events have been masked out, which is what we expect. The larger flares give rise to a higher GP error and are also more evident because they have a longer duration. However, even in the presence of large complex flares, we can still see the substantial curvature of the starspot modulation traced by the GP. In areas where small flares occur, the GP variance is significantly smaller.

\section{Continuous Flare Model}

\label{sec:flaretemp}

Flares share a common underlying formation mechanism, therefore a time dependent profile can be derived to model the observed flare morphology, as shown in D14. The median flare template can be described by an analytic function. To improve on the flare profile from D14, we use the convolution of a Gaussian and a double exponential to model the morphology of the flares as shown in \citet{Jackman2020DetectionSurvey}. This improves on the work of \citet{Gryciuk2017FlareCurves} who fit data of X-ray solar flares using the convolution of a Gaussian and a single exponential. Both approaches avoid the sharp flare peak (discontinuity) that is present in the D14 model. 

Mathematically, the flare profile, $f(t)$, is defined as:

\begin{equation}
    f(t) = \int_{-\infty}^{t} g(x)h(t-x)dx.
    \label{eqn1}
\end{equation}

The Gaussian term, $g(x)$, accounts for the impulsive heating that occurs during the rise phase of the flare, which has been used to model solar flares \citep{Aschwanden1998LOGISTICFLARES}, and takes the form:

\begin{equation}
    g(x) = A e ^{(\frac{-(x-B)^2}{C^2})}.
\end{equation}

Meanwhile, the double exponential, $h(x)$, accounts for the rapid and gradual cooling phases of the flare event that are described in D14:
\begin{equation}
    h(x) = F_1 e^{(-D_1 x)} + F_2 e^{(-D_2 x)}.
\end{equation}

By taking the convolution of these two functions we can account for the heating and cooling processes happening during each flare. Therefore, the updated flare template is based on a continuous function,

\begin{eqnarray}
f(t) = 
\frac{\sqrt{\pi}AC}{2} \times \cr
\Big( F_1  h(t,B,C,D_1) +F_2 h(t,B,C,D_2)\Big),
\label{eqn4}
\end{eqnarray}
where
\begin{eqnarray}
h(t,B,C,D) = e^{-Dt+(\frac{B}{C}+\frac{DC}{2})^2} \times \cr
erfc (\frac{B-t}{C}+\frac{DC}{2}), 
\end{eqnarray}

where erfc(t) is the complementary error function defined as 1-erf(t). The error function is commonly used in statistics and is defined as erf(t) = $\frac{2}{\sqrt{\pi}} \int exp (-s^2)ds$. It is available in the SciPy package (\texttt{scipy.special.erf}) for numerical evaluation.

The complete formula, $f(t)$, depends on the values of 8 parameters that help define the overall flare shape. These are:

\begin{itemize}
    \item []{$t$ = relative time,}
    \item []{$A$ = amplitude,}
    \item []{$B$ = position of the peak of the flare,}
    \item []{$C$ = Gaussian heating timescale,}
    \item []{$D_1$ = rapid cooling phase timescale,}
    \item []{$D_2$ = slow cooling phase timescale, and}
    \item []{$F_{2}$ = 1 - $F_{1}$, which describe the relative importance of the exponential cooling terms.}
\end{itemize}

We note the limits of integration in Equation \ref{eqn1} are different than those in \citet{Gryciuk2017FlareCurves}. Here we evaluate the model from $-\infty < x < t$ to correct for the fact that Gaussian functions are defined from $- \infty < x < \infty$. This mathematical correction also helped account for the divergent behaviour that was present with the previous parameterization implemented in \citet{Jackman2020DetectionSurvey}.

\subsection{Comparing Flare Model Parameterizations}

Fitting a continuous flare model to photometric observations gives us the ability to parameterize flare events. The convolution of a Gaussian and a single exponential template has been used to model flares and other explosive events such as supernovae \citep{Papadogiannakis2019R-bandFactory}. Many of these events have a characteristic FRED profile. A single exponential decay model has been used frequently, especially while searching for flares among large catalogs \citep[e.g.][]{Walkowicz2011White-lightData,Loyd2014FLUCTUATIONSOBSERVATIONS}. While this model accounts for the heating phase, it does not accurately model the decay phase of flare events. A two-phase cooling profile for flares on M-type stars was proposed by \citet{Andrews1965FlareStars}, which consisted of a sharp linear decline followed by an inverse square shape. The decay phase was later parameterized observationally by \citet{Hilton2011TheRate} with an initial linear decline and exponential profile. Spectroscopic analyses by \citet{Kowalski2013Time-resolvedSpectra} also found emission components that suggest there are to two distinct regions during the flare decay: one that cools more rapidly and another which cools slower.  More recently, D14 used a double exponential to model the thermal and non-thermal cooling processes happening during the decay of stellar flares.

\begin{figure}[!t]
    \centering
    \includegraphics[width=3.35in]{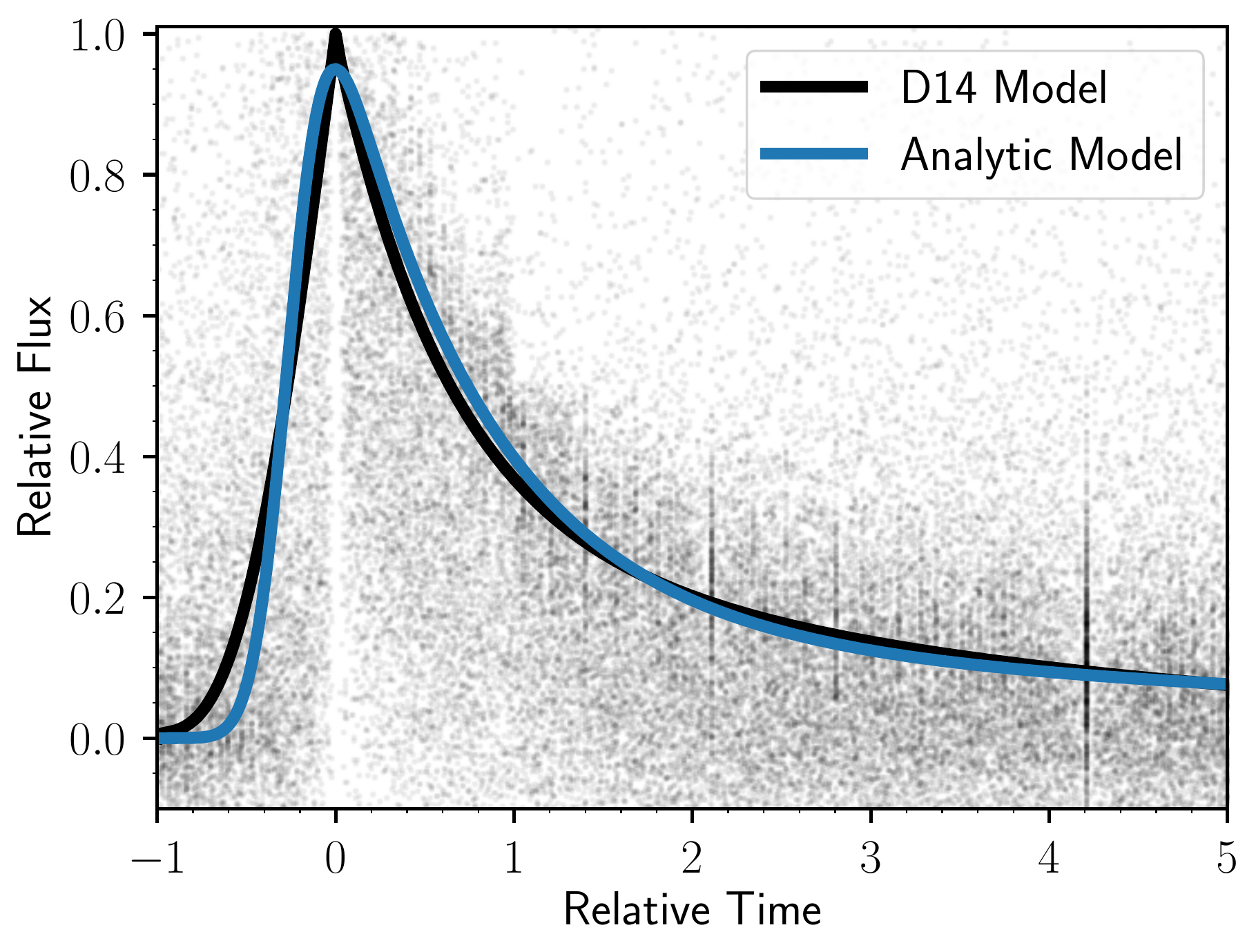}
    \caption{Comparison of the piece-wise flare template from D14 (black curve) and the initial version of our continuous flare model (blue curve). For reference, we show the full 885 flares sample stacked using the D14 procedure (grey points), which produces significant scatter in flux from forcing each flare to a peak of exactly 1. We also note the aliasing from improperly estimating the FWHM for short duration events (vertical bands) from the 1-minute cadence \Kepler data.}
    
    \label{fig:2_models}
\end{figure}

The convolution of a Gaussian and a double exponential has been shown to more accurately represent the heating and two-phase cooling processes that occur during flare events  \citep[e.g.][]{Jackman2018Ground-basedNGTS,Jackman2019DetectionSurvey}. Physically, we get a continuous and analytic model that allows us to parameterize classical, single-peaked flare events and later decompose complex events into a series of classical events as seen in D14. Figure \ref{fig:2_models} shows the comparison between the D14 flare template and our new template. In comparison to the D14 piece-wise model, the updated analytic model does a better job at modeling the peak of the flare events and does not pin each flare to a relative peak flux of one. By using the new parameterization, updated detrending, and new starspot modeling techniques the updated flare model greatly improves our ability to parameterize flare events.

\section{Stacking the Flares}

\label{sec:stacking}
Following the work of D14, we stacked all of the classical flare events onto a common time and flux axis to construct the flare template. The large sample of flares helps us achieve fine sampling. For instance, if the typical flare event is about 30 minutes long, we can get less than 1 second resolution by stacking hundreds of flares onto each other. By stacking the flares we get higher sensitivity to short timescale features.

We start by replicating the stacking procedure from D14. For each of the flares we first subtract off the starspot modulations from the mean GP, described in Section \ref{sec:gp regression}. Once the continuum is subtracted, we divide each flare by the maximum flux (peak) within each event. Each flare was therefore normalized to a relative flux scale between 0 (before and after the flare occurs) and 1 (peak flux). Each flare was also set to a relative timescale. To account for the rapid rise and decay phases, we linearly interpolated each flare to a time resolution that was 10 times higher to yield a more accurate value of the full time width at half the maximum of the flux (FWHM), also known as t$_{1/2}$ in previous studies \citep{Kowalski2013Time-resolvedSpectra}. This allowed us to reproduce the same flare stack that D14 created (see Figure \ref{fig:2_models}), which was based on three free parameters: peak time, scale time, and amplitude.

However, the stacking procedure used in D14 has a number of limitations. One major limitation is that all of the flares were pinned at an infinitesimal peak of exactly one. This means the peaks of flare events were systematically underestimated and it also increased the relative flux scatter in the stack, which can be seen in Figure \ref{fig:2_models}. By forcing all of the flares to align to a center time there is also an additional source of scatter added to the relative time. Furthermore, the stacking procedure used in D14, imprecisely estimated the FWHM of flares by using an arbitrary linear up-sampling of the light curve. This caused scatter in the relative time, which created aliasing or over-dense regions of the stack as seen in Figure \ref{fig:2_models} (vertical grey bands). These features are present in the D14 paper, but cannot be seen due to the logarithmic contour map used to present the data in D14 Figure 4. D14 also used local polynomials to detrend the starspots, which are dependent on the order of the polynomials and the flare masking. This technique also did not provide the associated starspot model uncertainties, which were therefore not incorporated into the D14 model.

\begin{figure*}
    \centering
    \includegraphics[width=7in]{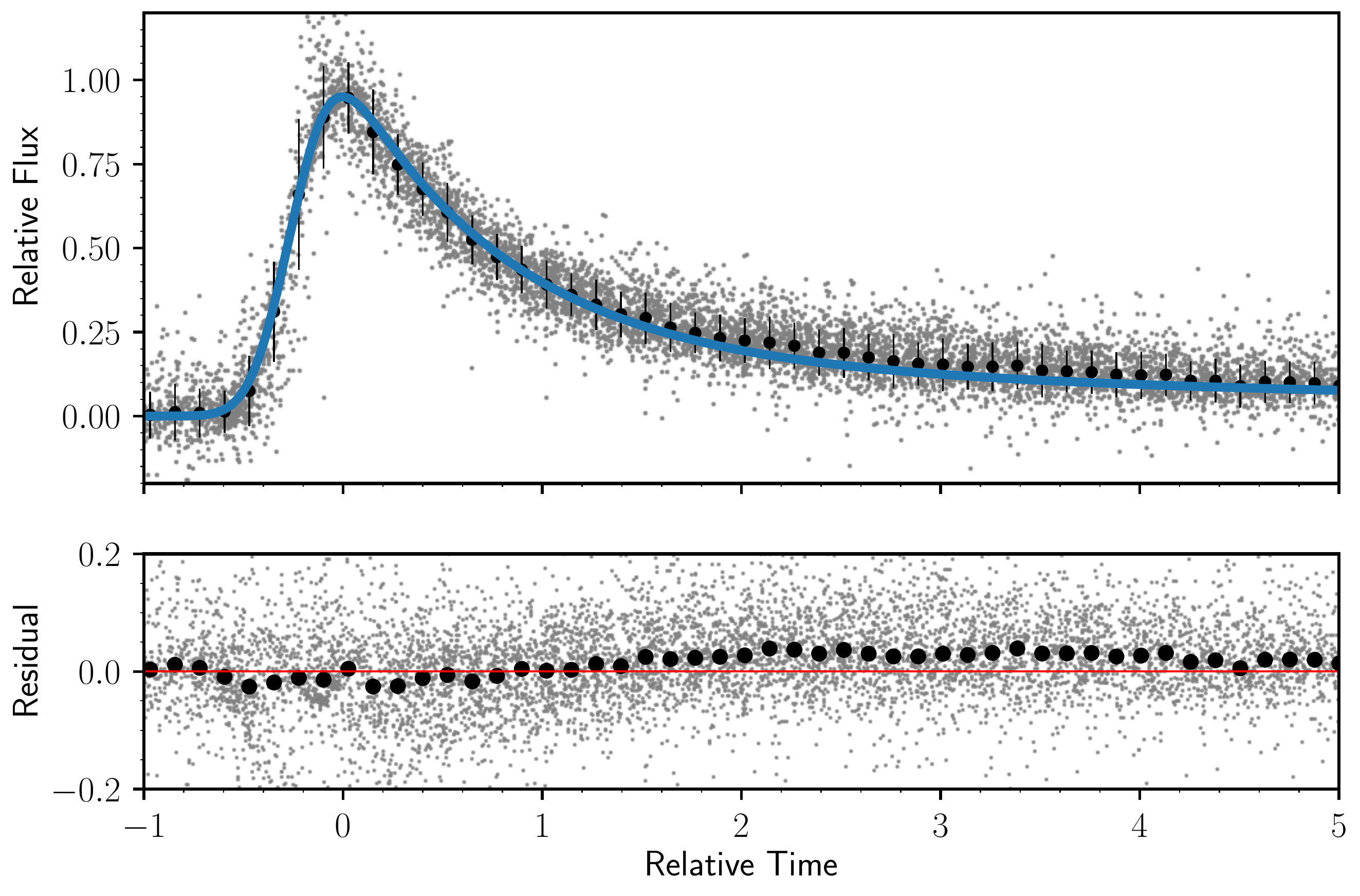}
    \caption{\textbf{Top:} The updated continuous analytical model (blue) overlaid onto the final vetted sample of 414 classical flares from the DR25 \Kepler data release. The solid black points are the binned median with the respective standard deviation of the points inside of each bin. The flares are overlaid using a new stacking procedure that is less sensitive to sampling effects and scales each flare to a relative time and amplitude. \textbf{Bottom:} The residuals of the model (grey) and the binned residuals of the model in black which are mostly uniform and flat. The structure in the decay phase of the flares is caused by a combination of uncertainties in the stop times of flares and the starspot detrending. The combination of using a GP to model the starspots + continuous model + vetted flare sample + new stacking procedure has produced a more robust flare template. }
    \label{fig:new_model}
\end{figure*}

\subsection{Improving the stacking procedure}
\label{sec:improved stack}

Here we present an updated stacking procedure that addresses the limitations of the D14 stacking procedure. First, we used a non-linear least squares optimization to fit the initial version of our flare model shown in Figure \ref{fig:2_models} to each of the 414 flares in our sample. We used the parameters of the individual flares from D14 to initialize our fits. Our least-squares fitting was weighted by the photometric and GP errors added in quadrature. We then conceptually used the same alignment procedure as D14 to stack the flares, scaling each flare by the fit amplitude and FWHM, and aligning each event by the center time. Note this center time may not exactly correspond to the observed peak. By using these fits to align the flares, we are able to produce a stack that is not dependent on the peak estimate from the light curve, as in D14.

In Figure \ref{fig:new_model}, we can see the updated model overlaid onto the stack of 414 classical flares that uses the new stacking procedure. In total, there are 13421 epochs of data represented among the stacked flares. We also show the binned median of the data (bins=200), and we can see the updated model traces the underlying shape of the flares. The stack of flares is much cleaner and has a reduced scatter in comparison to the stack of flares used in D14 (see Figure \ref{fig:2_models}). The updated model therefore uses an updated stacking procedure that is less sensitive to sampling effects.

To quantitatively compare the two stacking procedures, we first fit a rolling median to each of the flare stacks. Then we calculated and compared a reduced $\chi^2$ for each of the resulting rolling medians. For the D14 rolling median we compute a reduced $\chi^2$ of 16.5. Meanwhile we calculated a reduced $\chi^2$ of 10.8 for the rolling median of the latest stack of flares. The lower $\chi^2$ values that we calculated quantitatively demonstrate the improvements to the stacking procedure. 

We also tried other stacking procedures to test which approach would further improve the flare model. Specifically, we drew from work in the exoplanet community that uses cumulative distributions to understand the distributions of planet eccentricities \citep{Moorhead2011TheEccentricities}. However, this technique of using a cumulative distribution did not produce the correct center time and scale time alignments for flares. In comparison to the D14 model, we found that the updated stacking procedure used in this work both qualitatively and quantitatively improved the flare model.

\section{Fitting the New Model} 
\label{sec:fit}

\begin{deluxetable}{ccc}
\tabletypesize{\footnotesize}
\tablecolumns{6}
\tablewidth{0pt}
\tablecaption{ \label{table:flarevals}}
\tablehead{
\colhead{Parameter}&\colhead{Value} &\colhead{Uncertainty} }
\startdata
 A & 0.969 &  7\e{-3} \\
 B & -0.2513 & 4\e{-4} \\
 C & 0.2268 & 6\e-{4} \\
 D1 & 0.156 & 1\e{-3} \\
 D2 & 1.215 & 4\e{-3} \\
 F1 & 0.127 & 1\e{-3} \\ 
\enddata

\tablecomments{Best fit coefficients for Equation \ref{eqn4}, and their respective uncertainties from the Markov Chain Monte-Carlo analysis, which define the flare shape as shown in Figure \ref{fig:new_model}.}
\end{deluxetable}

\begin{figure*}[htbp!]
    \centering
    \includegraphics[width=7.1in]{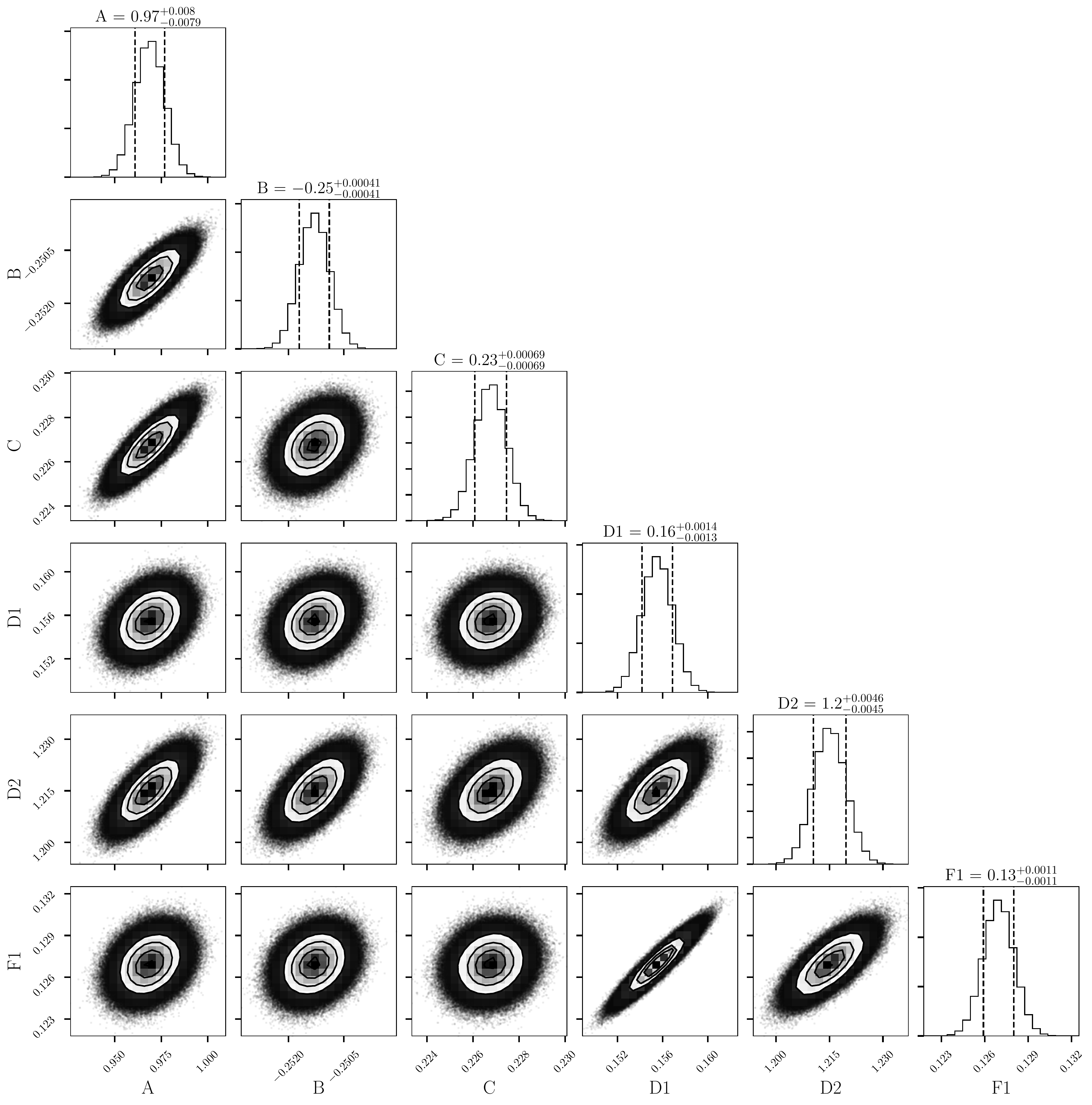}
    \caption{The results from the MCMC analysis showing the posterior probability distributions of each of the model parameters from Equation \ref{eqn4}. This figure was made using \texttt{corner.py} \citep{Foreman-Mackey2016Corner.py:Python}.}
    \label{fig:mcmc}
\end{figure*}


We use the python package \texttt{EMCEE} \citep{Foreman-Mackey2013EmceeHammer} to perform a Markov Chain Monte Carlo analysis, which fits the stacked flare sample. To initialize the walkers we used our flare fits (see Section \ref{sec:improved stack}). We ran \texttt{EMCEE} using 256 walkers, 30,000 steps and we discarded the first 10\% as burn in which we found was sufficient to reach convergence. The acceptance fraction was 0.516 with a mean auto correlation time of 65.01 steps. We used a $\chi^2$ test with a tight boundary ($D1 > 0$) for our log likelihood model and assume a flat prior on all parameters. In addition, we used both the photometric and GP uncertainties added in quadrature. We note that the photometric errors are the ones that primarily contribute to the scatter, and the GP errors don't exceed the photometric errors. In the bottom panel of Figure \ref{fig:new_model}, we include the residuals of the model and find that they are mostly uniformly scattered and flat. This tells us the data are well fit by our model. We see some structure in the residuals that correspond to the decay phase of the flare events. This structure is a result of uncertainties in the stop times of flares as well as uncertainties with the GP. During the end of the decay phases we reach comparable timescales with the starspots evolving and long tailed flares. Therefore, this is the regime where we are most affected by the GP detrending and the manual identification of the stop times of flares.

The best fit parameters from Equation \ref{eqn4} and their respective errors are presented in Table \ref{table:flarevals}. In Figure \ref{fig:mcmc}, we present the typical corner plot of the resulting MCMC analysis, which shows the posterior probability distributions for each of the six model parameters. This fit defines the new flare template shape. 

\begin{figure}[!t]
    \centering
    \includegraphics[width=3.35in]{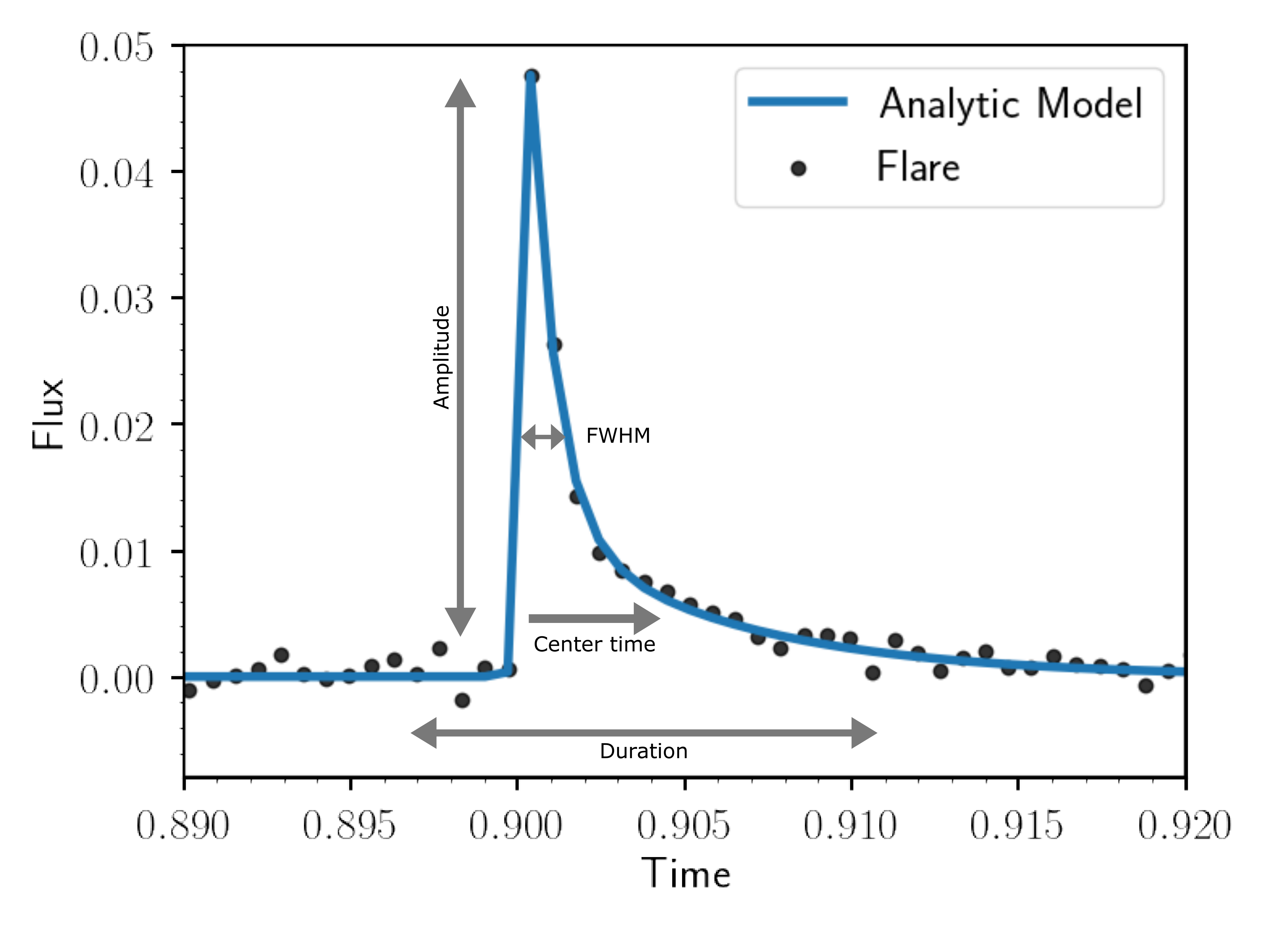}
    \caption{Example flare from the vetted set of \Kepler classical flares with the analytic model overlayed. We show the final flare model can be parameterized using the following three scaling parameters: amplitude, FWHM, also known as $t_{1/2}$ in \citet{Kowalski2013Time-resolvedSpectra}, and center time (which is similar to tpeak in D14).}
    \label{fig:annotated}
\end{figure}

As in D14, this new flare template can be applied to observations via the same three scaling parameters used in making our stacked flare sample in Figure \ref{fig:new_model}: center time, FWHM (also known as $t_{1/2}$ in \citet{Kowalski2013Time-resolvedSpectra}), and amplitude.
This is similar to the process used in scaling supernovae template to fit light curves \citep{Papadogiannakis2019R-bandFactory}. In Figure \ref{fig:annotated}, we show both an example of a classical flare profile that was modeled using our updated analytic flare template, as well as the three scaling parameters used to fit individual flares (center time, FHWM, and amplitude). The code for the updated flare model is made publicly available on GitHub \faGithub\href{https://github.com/lupitatovar/Llamaradas-Estelares}.


\subsection{Model Comparison}
\label{sec:6.1}

\begin{figure}[!t]
    \centering
    \includegraphics[width=3.35in]{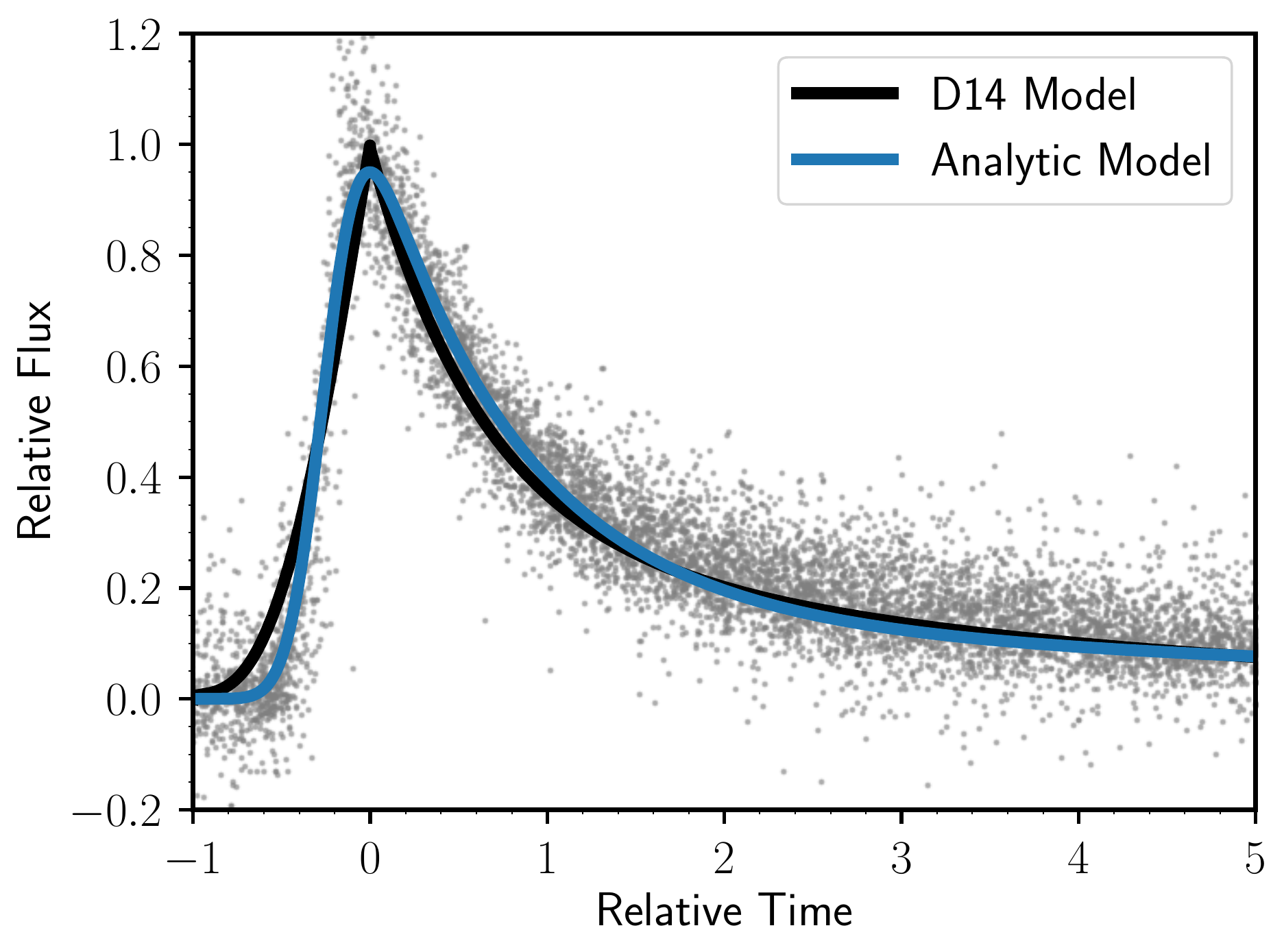}
    \caption{Comparison of the D14 and updated analytic flare templates overlayed onto the new stacked flares data set (grey points). The scatter in the data is from nearby, low energy flares. In black, is the piece-wise model from \citet{Davenport201411243} and in blue one is the new analytic template that uses the convolution of a Gaussian and a double exponential. The updated analytic model is continuous and more accurately describes the peak of the flare events.}
    \label{fig:2_models_new}
\end{figure}

The updated analytic flare model is both qualitatively and quantitatively more robust than the previous model presented in D14. In Figure \ref{fig:2_models_new}, we can see the two flare models overlayed onto the vetted, 414 classical flares that were stacked using the new stacking procedure described in Section \ref{sec:stacking}. Qualitatively, we can see the scatter from the stacked flares is both reduced and more uniform in comparison to the stacked flares shown in Figure \ref{fig:2_models}. With the new stacking procedure we also account for the aliases that were present in the D14 stacking procedure that were a result of the alignment and scaling procedure used. 

To quantitatively compare the two models, we calculated the reduced $\chi^2$ for each of the stacking procedures that were used to derive the respective flare models. Using the D14 model and updated stacking procedure described in Section \ref{sec:improved stack}, we calculated a reduced $\chi^2$ of 13.9. Meanwhile, the reduced $\chi^2$ that uses the updated model derived in Section \ref{sec:fit} and shown in Figure \ref{fig:new_model} is 9.1. This is a lower value in comparison to D14, showing the new model is a better fit to the stack. In addition, we fit the vetted sample of 414 \Kepler classical flares with both D14 and the new model. This allowed us to compare the changes between the resulting model fits for the same set of flares. We calculated the $\chi^2$ of both models for every individual flare as a function of log flare energy. At higher flare energies ($log E > 31$) we find the updated model shows a lower $\chi^2$, indicating the new model provides a better fit for higher energy flares. This is likely due to decreased resolution for lower energy flares. Overall, the updated model presented in this work has a lower $\chi^2$ for an individual flare in comparison to the D14 model. 

\subsection{Updated Flare Properties}
\label{sec:6.2}

\begin{figure}[t!]
    \centering
    \includegraphics[width=3.35in]{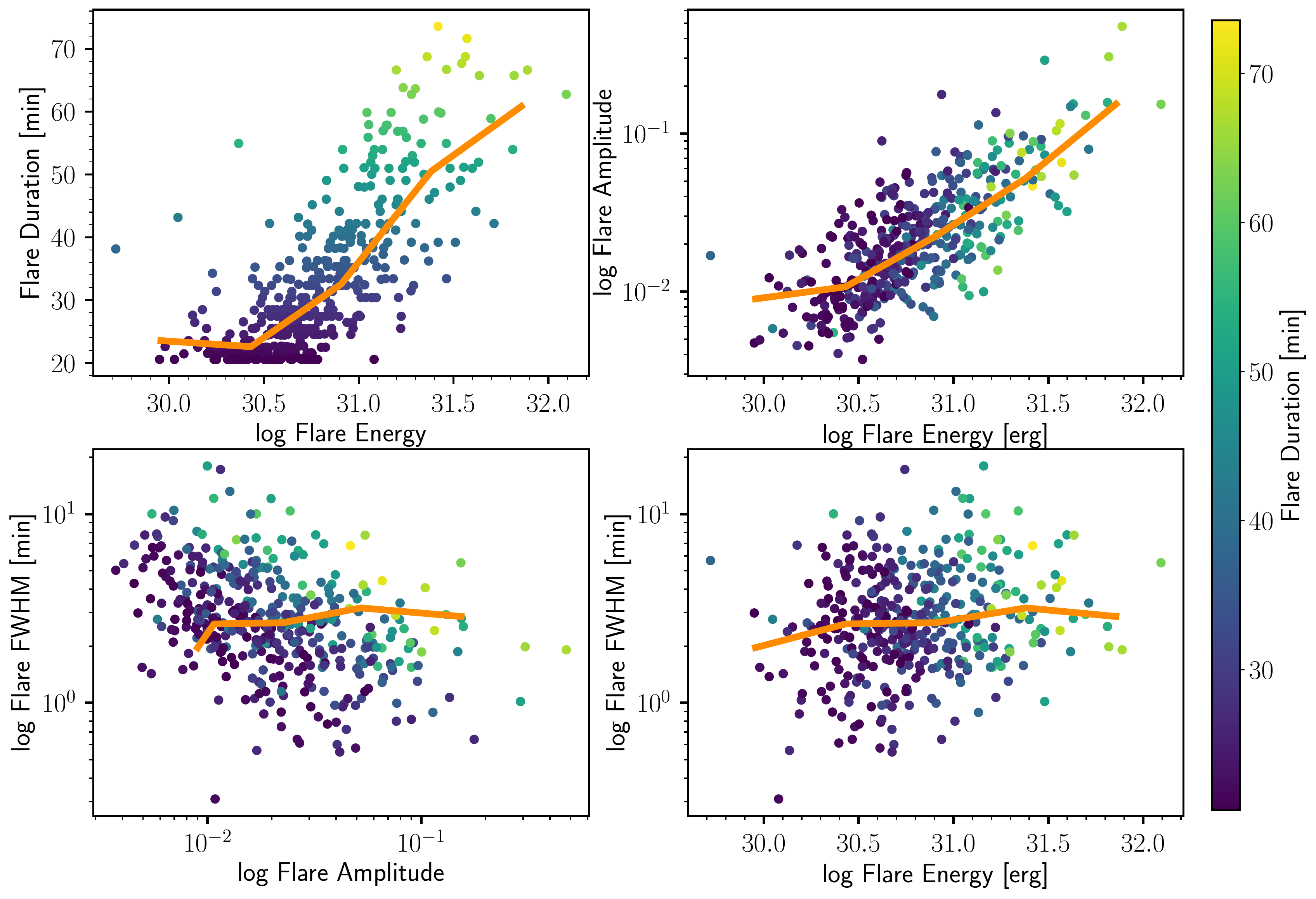}
    \caption{\textbf{Top:} The top two panels follow the correlation we expect. Flares with higher energies occur over longer timescales and also have higher amplitudes. Meanwhile, lower energy flares occur over shorter timescales and have shorter amplitudes. \textbf{Bottom:} There is significant scatter among the FWHM plots which shows that there are flares that are both tall and narrow as well as short and wide. The combination of FWHM and flare amplitudes allow us to characterize the flare profile. The orange lines represent the binned medians for each of the datasets. The color gradient shows flares with different durations (yellow = longer flares and purple = shorter flares).}
    \label{fig:params}
\end{figure}


\begin{figure}[t!]
    \centering
    \includegraphics[width=3.35in]{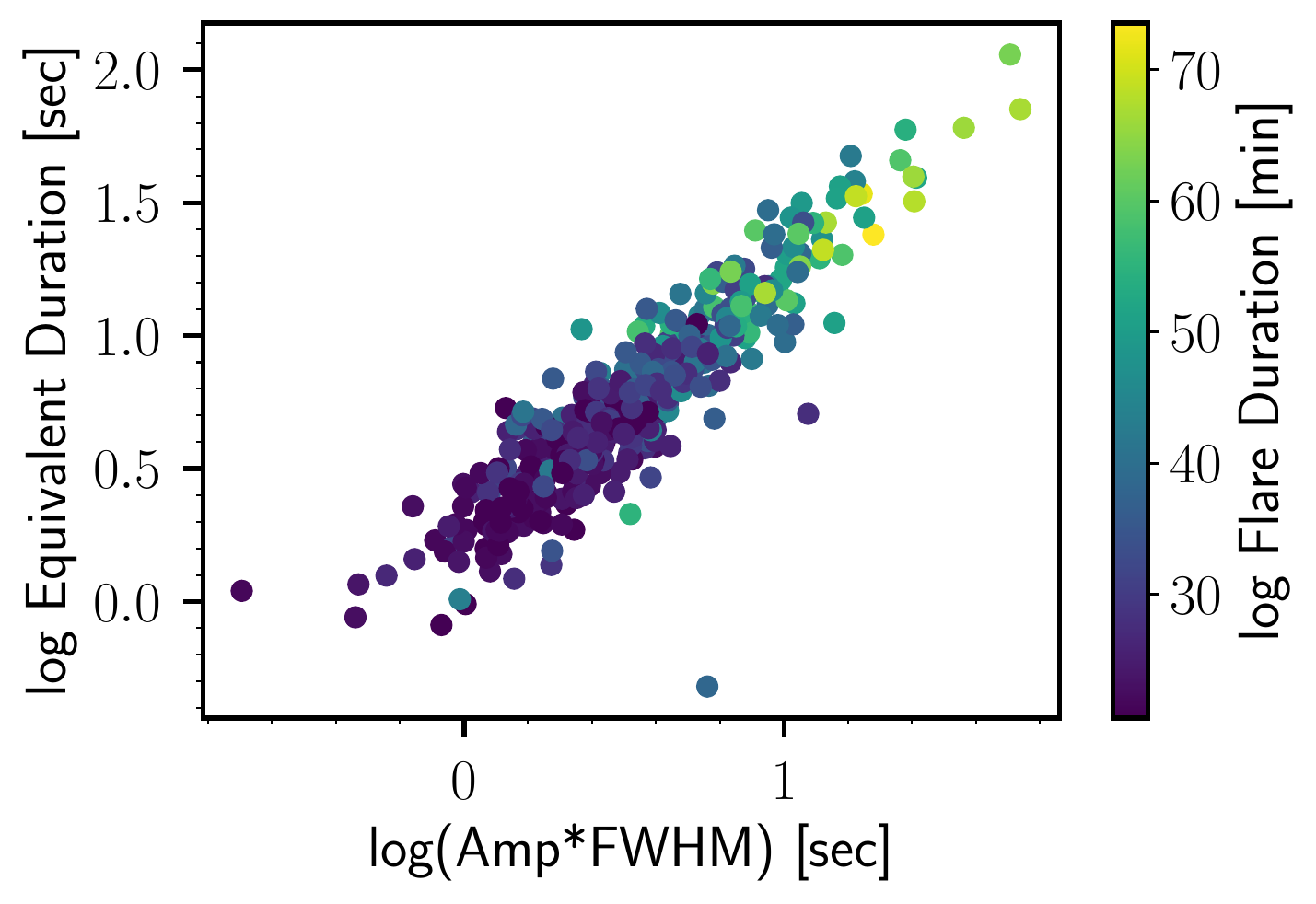}
    \caption{Here we show the product of FWHM and amplitude versus the equivalent duration of flare events, both in units of seconds. The data points are color coded by time durations in units of minutes, where lighter colors correspond to longer flare events and darker colors correspond to shorter flare events. The tight correlation between these parameters indicates taht the combination of flare amplitude and FWHM are sufficient to characterize the flare event profile with our model. }
    \label{fig:eve}
\end{figure}

We use the flare fits described in Section \ref{sec:6.1} of the 414 classical flare events from the \Kepler data to explore the relationships among the various flare parameters. We include these flare fits in Table \ref{table:flare fits} so future studies may use properties of real flare events as inputs for various simulations. In Figure \ref{fig:params}, we show the relationship between the fit flare parameters: amplitude, full time width at half of the flux maximum (FWHM), and duration as a function of the event energy. The equivalent duration is computed by integrating the fractional flux under each flare  \citep[]{Gershberg1972,Hunt-Walker2012MOSTLeo1} and is used to measure the flux event energies. The correlations between flare energy and flare duration are consistent with what we expect: higher energy events occur over longer timescales, while short duration flares typically have lower energies. Similarly, higher energy flares have larger amplitudes. The same trend is true when we consider FWHM as a function of flare energy, however there is a larger scatter in this correlation. 

Previous studies have explored the physical interpretation of the correlation between flare energy and duration. For instance, \citet{Maehara21} carried out time-resolved photometry and spectroscopy of the M-type star YZ CMi and found that the duration of flares showed a positive correlation with the flare energy. Specifically, they find the duration of flares increases with energy as $\tau_{flare} \propto E_{flare}^{0.21 \pm 0.04}$. However, this is a lower correlation than was found for G-type stars, which suggests a higher coronal magnetic field strength around active M-type stars like YZ CMi and GJ 1243 \citep{Maehara21}. This timescale versus energy relation is consistent with our FWHM versus energy plot in Figure \ref{fig:params}, which is expected since GJ 1243 and YZ CMi are of similar mass. We note the timescales are not exactly the same because we use flare FWHM and \citet{Maehara21} uses e-folding time. 

To further show how these parameters characterize the flare event profile, in Figure \ref{fig:eve} we consider the correlation between the equivalent duration and the product of the FWHM and amplitude. This product effectively gives an equivalent duration for the impulsive phase of the flare (see Figure \ref{fig:annotated}), which D14 showed only encompasses about one third of the total event energy.
There is a tight correlation of this product with the flare total energy, which parallels a 1-to-1 trend in Figure \ref{fig:eve}.
This demonstrates that while there can be a large scatter in the individual flare properties (e.g. Figure \ref{fig:params}), the total event can be robustly described by these two impulsive properties (FWHM, Amp), which we use to scale our flare template when fitting actual events. 

\section{Applications}
\label{sec:apps}

\begin{figure*}
    \centering
    \includegraphics[width=7in]{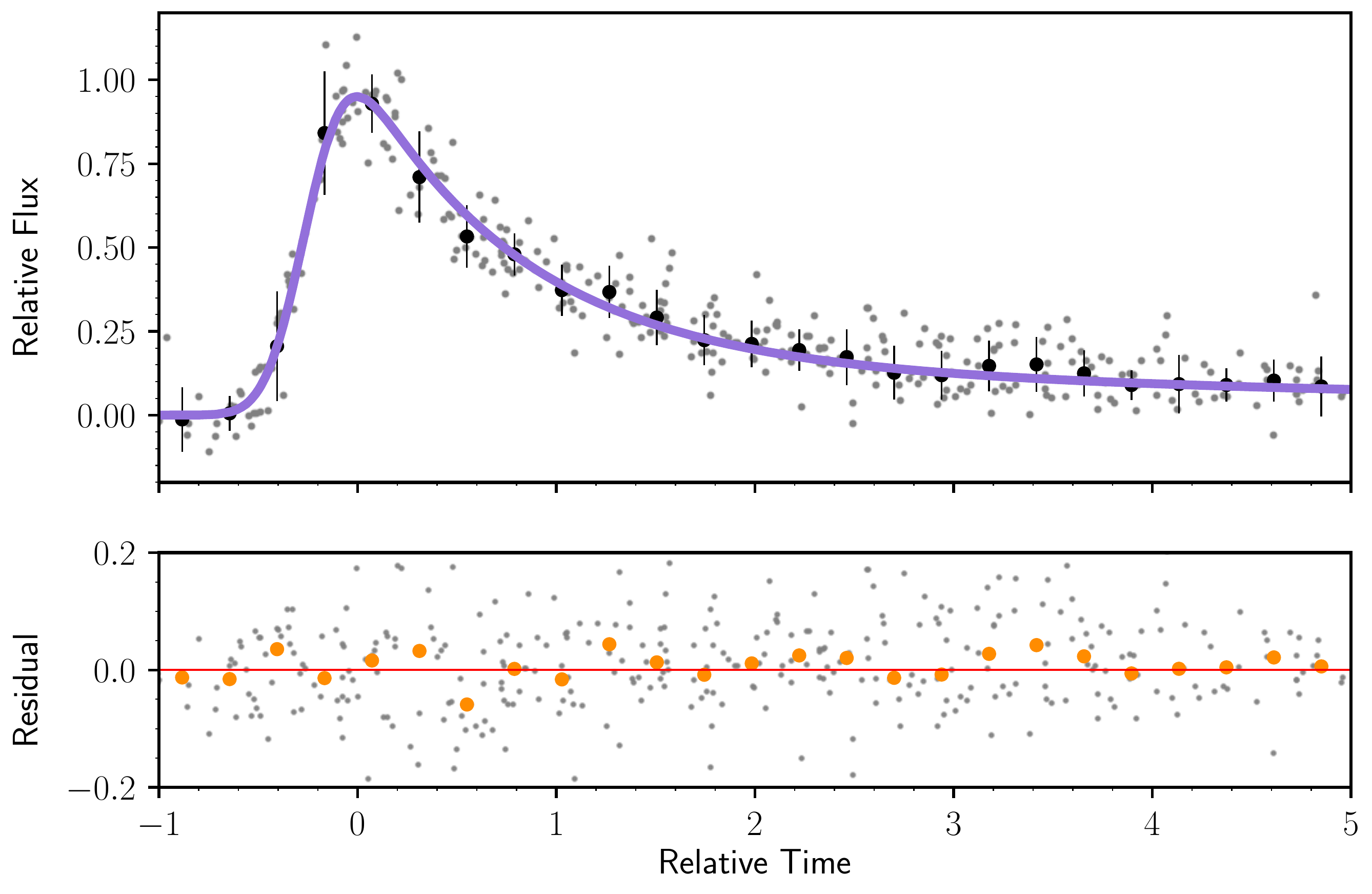}
    \caption{We test the GP modeling, stacking procedure and flare identification on \TESS data. \textbf{Top:} Overlay of the updated flare template (purple) onto 25 classical flares from sectors 14 and 15 of GJ 1243 \TESS data (grey) that are all scaled to a relative time and amplitude. The black points are the binned median of the data with the respective standard deviation of the points inside of each bin. \textbf{Bottom:} The residuals of the model (grey) and the binned median of the residuals (orange). The scatter is fairly uniform and the residuals are low suggesting the data is well fit by the model. The flare template can be used to model flares from different datasets and observations at different cadences.}
    \label{fig:tess_stack}
\end{figure*}

To test our new analytic flare model using a different data set, we turned to the Transiting Exoplanet Survey Satellite \citep{Ricker2014TransitingTESS}. \TESS recently revisited the \Kepler field, which included GJ 1243 (TIC 273589987) and provided us with 50 days of new short-cadence (2-minute) observations from Sectors 14 and 15 \citep{Davenport2020101243}. \TESS has provided the most detailed light curve for this star since the end of the original \Kepler mission, and has observations that are at longer and redder wavelengths relative to the \Kepler bandpass. 
By convolving a 10,000 K blackbody curve with each of the filters, \citet{Davenport2020101243} showed that within typical uncertainties for distance and flux calibrations, \TESS and \Kepler are well suited for comparison since they have similar flare energy yields. They also found that when the flare frequency distributions were modeled for each data set the flare activity remained unchanged from the \TESS to \Kepler epochs \citep{Davenport2020101243}. This provided us with the ideal opportunity to test our new flare model on a data set that had both a different cadence and wavelength coverage for the same star.

\subsection{GJ 1243 \TESS Data}

\begin{figure*}[htbp!]
    \centerline{\includegraphics[width=7in]{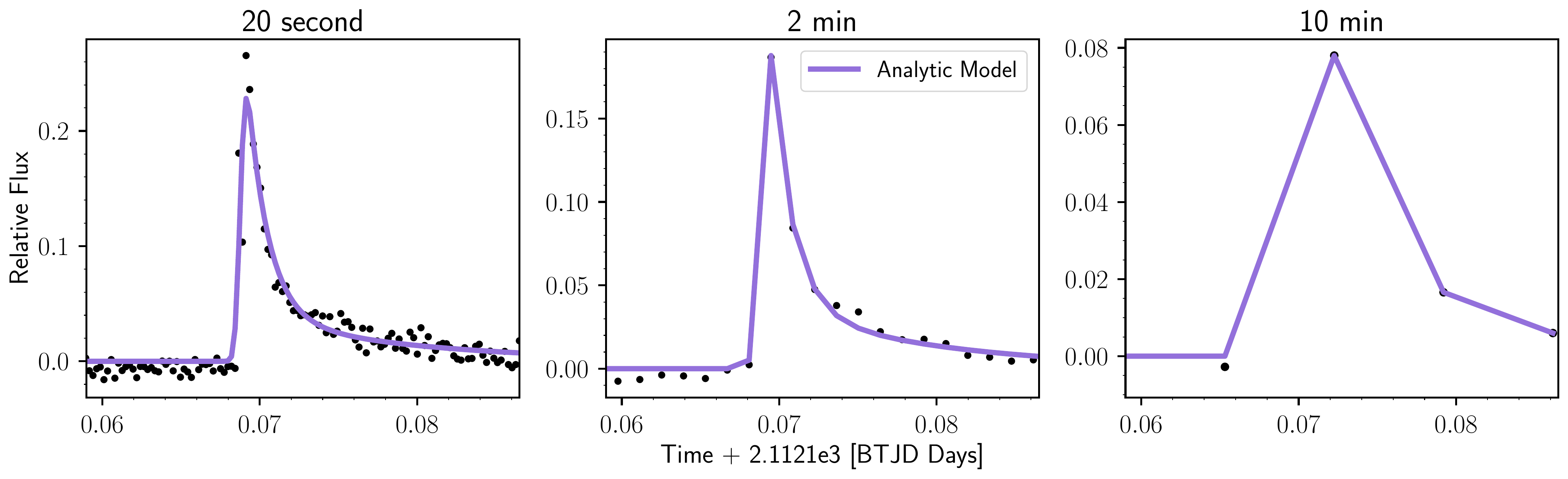}}
    \caption{We use the new analytic flare template to model one of the flare events occurring on the M3.5 star (TIC 197829751) that was observed by \TESS. We compare how our model works at different time resolutions. On the far \textbf{Left:} we have 20 second data, in the \textbf{Center:} we have 2 minute data, and on the \textbf{Right:} is the 10 minute data for the same target. Note the increased amount of structure that is revealed in the 20-second flare data vs the 10-minute data. The higher resolution highlights some secondary features in the cooling phase of this particular flare that are not as apparent in the 2-min or 10-min data, hence the different model fits. A table of the best fit values for each cadence mode can be found in Table \ref{table:tess flare}.}
    \centering
    \label{fig:exs}
\end{figure*}

We use the set of 133 flares (classical and complex) from GJ 1243 that were identified using the FBEYE tool by \citet{Davenport2020101243} in \TESS sectors 14 and 15.  We compared the FBEYE catalog to \texttt{stella}, an algorithm that uses a convolutional neural network (CNN) to find flares \citep{Feinstein2020FlareData}. To initialize \texttt{stella}, we set a flare finding threshold of 0.75, which limits what light curve features get classified as flares versus non-flares. We ensemble the 10 CNN training models provided in \citet{Feinstein2020FlareData} and average over the predictions of each of the training models. Ensembling provides a more robust flare classification, which reduces false positives and provides a higher confidence in the true positives. In the end, \texttt{stella} successfully recovers 75 flares from the same light curve. We note that GJ 1243 is near the stellar rotation period limit of \texttt{stella} (0.59 days), which can be a reason for the discrepancy in the total number of flares that were identified by each flare-finding technique. However, \texttt{stella} serves as a fast and reliable tool for finding flare events in the \TESS two minute light curves. Further by-eye analysis is needed to determine which flares are complex versus classical from the \texttt{stella} catalog. For this data set, we identified 25 classical flares that were within the 20-75 minute duration range, and we used these flares to test our model.

As in the procedure used for the \Kepler data, we masked out the flares and used a GP to model the starspot modulations. We used the same initial parameters and rotation period as described in Section \ref{sec:gp}. Once we had modeled the starspot variability, we stacked the classical flares onto a common relative time and flux space using our updated stacking procedure described in Section \ref{sec:stacking}. We were then able to overlay our analytical model onto the stack of \TESS classical flares. In Figure \ref{fig:tess_stack}, we see the results of our analytic flare template from Section \ref{sec:fit} overlayed onto the 25 stacked \TESS flares. We also show the binned median of the data. Similar to the case with \Kepler, we find that the residuals are mostly uniform and flat. Again, we find the photometric errors primarily outweigh the GP errors. Given the small sample of \TESS classical flares (25) we do not present a \TESS specific model for this work. Instead, we apply our existing model to the \TESS data to show its versatility. We find that our flare model can be used to model the morphology of white-light flare events from other data sets with differing observation cadences (e.g. 1-minute versus 2-minute).

\subsection{Flares at Different Cadences}

\TESS also provides 20 second observations for a subset of stars. This presented an opportunity to study how flares change across observational cadences. For instance, \citet{Shibayama2013SUPERFLARESSUPERFLARES} showed that \Kepler super flares at 1-minute and 30-minute cadence have similar measured flare energies within the errors. Here we select a low mass flare star (similar to GJ 1243) from \TESS and use our flare template to model one of the highest signal-to-noise, classical flare events in the light curve. The target (TIC 197829751) is an M3.5 star \citep{Schneider2019ACRONYMMembers} with a rotation period of about 3.1 days. \TESS observed the star at 20-second, 2-minute, and 10-minute cadence modes in Sector 29 (see Figure \ref{fig:exs}). In the 20-second cadence data we are able to see the finer structure and complexity of the flare event that is revealed. Meanwhile, the 2-minute data does not show as much detail, especially in the decay phase of the flare event. The 10-minute data shows even less structure of the flare than the 2-minute data. It is interesting to note that the flare event complexity revealed in the 20 second cadence \TESS data might provide further examples of late phase EUV brightening \citep[e.g.][]{Liu2015EXTREMELYFLARES,Chen2020Extreme-ultravioletFlares} and also quasi-periodic pulsations (QPPs) \citep[e.g.][]{Pugh2016StatisticalKepler,Howard2021NoFlares}.

Higher time resolution observations allow us to both detect smaller flares and understand the complexity of the events. Longer cadence data misses short flares or confuses them for classical events instead of resolving the multi-peak, complex structure. In Figure \ref{fig:exs}, we use our flare template to model the same flare event at three different cadences. We provide the bet fit parameters for each of the three cadence modes in Table \ref{table:tess flare}. At the 2-minute and 10-minute cadence modes our template is able to describe the morphology of the flare. However, at the 20-second cadence mode it becomes clear that even classical flares reveal complex behavior given high enough time resolution \citep{Howard2021NoFlares}. Overall, this example highlights the flexibility of our flare model when used with data taken at various cadences. 


\begin{deluxetable}{cccc}
\tabletypesize{\footnotesize}
\tablecolumns{6}
\tablewidth{0pt}
\tablecaption{TIC 197829751 Flare Properties \label{table:tess flare}}
\tablehead{
\colhead{Cadence}&\colhead{$t_{peak}$} &\colhead{FWHM}  & \colhead{Amp} }
\startdata
20 second & 0.0692 & 0.0015 & 0.2408 \\
2 minute & 0.0691 & 0.0016 & 0.2363\\
10 minute & 0.0702 & 0.0011 & 0.3789 \\
\enddata
\tablecomments{Best fit coefficients for the example M3.5 \TESS flare at three different cadence modes (see Figure \ref{fig:exs}).}
\end{deluxetable}

\section{Discussion and Conclusions}
\label{sec:discuss}


\begin{deluxetable*}{ccccccc}
\tabletypesize{\footnotesize}
\tablecolumns{7}
\tablewidth{0pt}
\tablecaption{ Classical Flare Properties \label{table:flare fits}}
\tablehead{
\colhead{$t_{peak}$}  &  \colhead{FWHM} & \colhead{Amp}  & \colhead{$t_{peak}$ Error} &  \colhead{FWHM Error} & \colhead{Amp Error} & \colhead{Source} \\  
\colhead{(days)} & \colhead{(days)} & \colhead{(relative)} & \colhead{(days)} & \colhead{(days)} & \colhead{(relative)}}

\startdata
539.6503088	& 0.00153389 & 0.008934698 & 6.49E-05 & 0.000213685	& 0.000716695 & \Kepler \\ 
1685.732989	& 0.005316529 &	0.011036421 & 0.000218928 & 0.000694793 &	0.000953418	& \TESS
\enddata 
\tablecomments{Flare properties of 414 classical \Kepler flares and 25 classical \TESS flares. These are the results of our model fits as described in Section \ref{sec:6.2}. This table is available in its entirety in machine-readable form in the online journal. A portion is shown here for guidance regarding its form and content.}
\end{deluxetable*}

We have developed an updated analytic flare template to describe the morphology of white-light flares in precise, space-based photometry. The analytical model is an update to the piece-wise model generated in \citet{Davenport201411243}. Using a combination of improved \Kepler light curve processing, an improved flare parameterization from \citet{Jackman2018Ground-basedNGTS}, and new detrending techniques to account for background starspot variability, we generated an analytic and continuous flare template that can be used to model the white-light flare events on active stars. The flare model shape is described by a function that uses the convolution of a Gaussian and a double exponential, with six coefficients defining the flare morphology. We used a total of 414 unique classical flares from 11 months of GJ 1243 \Kepler data to derive the model. Our updated stacking procedure avoids using the peak flare time or height as parameters as these are sensitive to sampling effects and causes us to systematically underestimate the peak of the flares. Instead we used a center time that leverages the entire flare profile data, which reduced the uncertainty in the stacking procedure. By stacking hundreds of flares together we were able to get fine sampling (e.g less than one second effective resolution) and higher sensitivity to short timescale features (e.g. a smooth turn over at the peak of flares). The final analytic model can be used to fit individual flare events using the following three scaling parameters: amplitude, center time, and FWHM, which is also known as t$_{1/2}$ in previous studies \citep{Kowalski2013Time-resolvedSpectra}. The model can also be used to model white-light flare events on other stars and with different datasets.

We also studied the morphology of GJ 1243 flares in both the \Kepler and \TESS datasets. The data set of 133 flare events detected by \TESS allowed us to test our updated analytic flare template using a new data set. We see more scatter in the 2-minute \TESS sample in comparison to the 1-minute \Kepler data due to cadence and signal-to-noise differences but find our updated flare template is able to model the flares from a different dataset. Our model can be applied to data from different photometric observations at different cadences, which will prove useful when coupled to other stellar variability and transit modeling algorithms. 

In addition to GJ 1243, we analyzed another low mass star (TIC 197829751) that was observed by \TESS at three different cadences. We were able to use our analytic flare template to model one of the classical flares on the star. We find our model works well at modeling the flares on other active stars. It also can be used to model flares from different datasets and observation cadences. In the 20-second cadence observations, finer flare structure is revealed within a single flare event, whereas in the 10-minute cadence observations the data are more sparse, making it more difficult to see the entirety of the flare shape. However, in all cases our model was able to characterize the underlying flare shape.

Future studies will be able to use the analytical template to model complex (multi-peak) flare events. In the updated flare sample we classified 379 complex flare events that have not yet been modeled with our flare template. Similar to the work in D14, our template can be used as a model to decompose complex flares. This is under the assumption that complex events can be described as the superposition of many classical flares. By linearly adding the models one could use a series of analytic templates to describe the multi-peak flare events. Modeling complex flares will be important especially as we get shorter cadence observations (e.g \TESS 20 second targets) that are capable of resolving additional complexity present during the flare event. For instance, \citet{Howard2021NoFlares} sampled 226 flare stars using \TESS 20 second data and found 49 candidates experienced quasi periodic pulsations (QPPs) and another 17\% of the sample showed complex flare morphology. By decomposing the complex flares using the updated model we can compare and model the QPPs and complex structure revealed from new observations.

Beyond flare studies, our analytic flare template will also be a powerful tool for modeling transits in the presence of flares. Currently, many planet detection algorithms account for transits and flares separately \citep[e.g.][]{Luger2017ATRAPPIST-1} or the flare is simply masked out \citep[e.g.][]{Plavchan2020AMicroscopii}, which increases the uncertainty in the transit profile. More recent studies have simultaneously modeled the stellar activity within the planet search algorithm \citep{Gilbert2021NoData}, however, the flare template used in the analysis is from D14, which is improved upon in this work. Our new flare template can be incorporated into existing detection and characterization tools \citep[e.g.][]{Gunther2021Allesfitter:Velocity,Gilbert2022FlaresObservations}. By combining the stellar flare and starspot analysis methods described here with transit models we will be able to both refine existing star-planet parameters and search for transiting exoplanets around active stars that have not yet been detected.


\acknowledgments

We thank Benjamin Montet, Emily Levesque, and the anonymous reviewer for their insightful suggestions, which greatly improved this manuscript. 

The authors acknowledge that this research was conducted on the traditional lands of the Coast Salish peoples, the lands which touch the shared waters of all tribes and bands within the Duwamish, Suquamish, Tulalip and Muckleshoot nations. We honor with gratitude the land itself and the tribes.

GTM acknowledges support from the National Science Foundation Graduate Research Fellowship Program under Grant No. DGE-1762114. Any opinions, findings, and conclusions or recommendations expressed in this material are those of the author(s) and do not necessarily reflect the views of the National Science Foundation.

JRAD acknowledges support from the DIRAC Institute in the Department of Astronomy at the University of Washington. The DIRAC Institute is supported through generous gifts from the Charles and Lisa Simonyi Fund for Arts and Sciences, and the Washington Research Foundation.

This research was supported by the National Aeronautics and Space Administration (NASA) under grant number 80NSSC19K0375 from the TESS Cycle 1 Guest Investigator Program, 
and grant number 80NSSC18K1660 issued through the NNH17ZDA001N Astrophysics Data Analysis Program (ADAP).

This research has made use of the NASA Exoplanet Archive, which is operated by the California Institute of Technology, under contract with the National Aeronautics and Space Administration under the Exoplanet Exploration Program.

\software{Python, IPython \citep{Perez2007IPython:Computing}, 
NumPy \citep{Harris2020ArrayNumPy}, 
Matplotlib \citep{Hunter2007Matplotlib:Environment}, 
SciPy \citep{Virtanen2020SciPyPython}, 
Pandas \citep{Reback2021Pandas-dev/pandas:1.2.3}, 
Astropy \citep{Robitaille2013Astropy:Astronomy},
Lightkurve \citep{LightkurveCollaboration2018Lightkurve:Python},
celerite \citep{Foreman-Mackey2018ScalableCelerite},
\texttt{EMCEE} \citep{Foreman-Mackey2013EmceeHammer},
\texttt{stella} \citep{Feinstein2020FlareData},
Llamaradas Estelares \citep{gtm_fm}}


\end{document}